\documentclass[a4paper,aps,prl,superscriptaddress,column,reprint]{revtex4-1} 
\usepackage{graphicx}
\usepackage[small,justification=justified]{caption} 
\usepackage{braket}
\usepackage[utf8]{inputenc}
\usepackage{hyperref}
\usepackage{amsmath} 
\usepackage{amssymb}  
\usepackage{amstext}  
\usepackage{graphicx}
\usepackage{amssymb}
\usepackage{color}
\usepackage{subcaption}
\usepackage{natbib}

\begin{document}

\title{Improving a solid-state qubit through an engineered mesoscopic environment}

\author{G. \'Ethier-Majcher}
\thanks{These authors contributed equally to this work}
\affiliation{Cavendish Laboratory, University of Cambridge, JJ Thomson Avenue, Cambridge, CB3 0HE, UK}

\author{D. Gangloff}
\thanks{These authors contributed equally to this work}
\affiliation{Cavendish Laboratory, University of Cambridge, JJ Thomson Avenue, Cambridge, CB3 0HE, UK}

\author{R. Stockill}
\affiliation{Cavendish Laboratory, University of Cambridge, JJ Thomson Avenue, Cambridge, CB3 0HE, UK}

\author{E. Clarke}
\affiliation{EPSRC National Centre for III-V Technologies, University of Sheffield, Sheffield, S1 3JD, UK}

\author{M. Hugues}
\affiliation{Universit\'e C\^ote d'Azur, CNRS, CRHEA, rue Bernard Gregory, 06560 Valbonne, France}

\author{C. Le Gall}
\affiliation{Cavendish Laboratory, University of Cambridge, JJ Thomson Avenue, Cambridge, CB3 0HE, UK}

\author{M. Atat{\"u}re}
\email[Electronic address: ]{ma424@cam.ac.uk}
\affiliation{Cavendish Laboratory, University of Cambridge, JJ Thomson Avenue, Cambridge, CB3 0HE, UK}

\begin{abstract}
A controlled quantum system can alter its environment by feedback, leading to reduced-entropy states of the environment and to improved system coherence. Here, using a quantum dot electron spin as control and probe, we prepare the quantum dot nuclei under the feedback of coherent population trapping and measure the evolution from a thermal to a reduced-entropy state, with the immediate consequence of extended qubit coherence. Via Ramsey interferometry on the electron spin, we directly access the nuclear distribution following its preparation, and measure the emergence and decay of correlations within the nuclear ensemble. Under optimal feedback, the inhomogeneous dephasing time of the electron, $T_2^*$, is extended by an order of magnitude to $39$~ns. Our results can be readily exploited in quantum information protocols utilizing spin-photon entanglement, and represent a step towards creating quantum many-body states in a mesoscopic nuclear spin ensemble.

\end{abstract}

\maketitle

The interaction between a qubit and its mesoscopic environment offers the opportunity to access and control the ensemble properties of this environment. In turn, tailoring the environment improves qubit performance and can lead to non-trivial collective states. Significant steps towards such control have been taken in systems including nitrogen-vacancy centers coupled to $^{13}$C spins in diamond \cite{Togan2011}, superconducting qubits coupled to a microwave reservoir \cite{Murch2012}, and spins in electrostatically defined \cite{Reilly2008,Vink2009,Bluhm2010} and self-assembled \cite{xu2009} quantum dots (QDs) coupled to the host nuclei. In InGaAs QDs, the hyperfine interaction permits spin-flip processes to occur between a confined electron and the QD nuclei. Optical pumping of the electron spin induces a directional flipping of nuclear spins leading to a net polarisation buildup \cite{Urbaszek2013}. The resulting effective magnetic (Overhauser) field can be as strong as 7~T \cite{Tartakovskii2007}, leading to significant shifts of the electron-spin energy levels \cite{Eble2006,Urbaszek2007,Tartakovskii2007,Maletinsky2007a}. In contrast to other systems, the polarisation of this isolated mesoscopic ensemble can persist for hours \cite{Latta2011}. Coupling the electronic energy shifts to the optical pumping rate closes a feedback loop \cite{Greilich2007,Hogele2012,Latta2009,Ladd2010} that allows for selection of the degree of nuclear spin polarisation.

  \begin{figure*}[hbpt]
  	\begin{center}
 	\includegraphics[width=16cm]{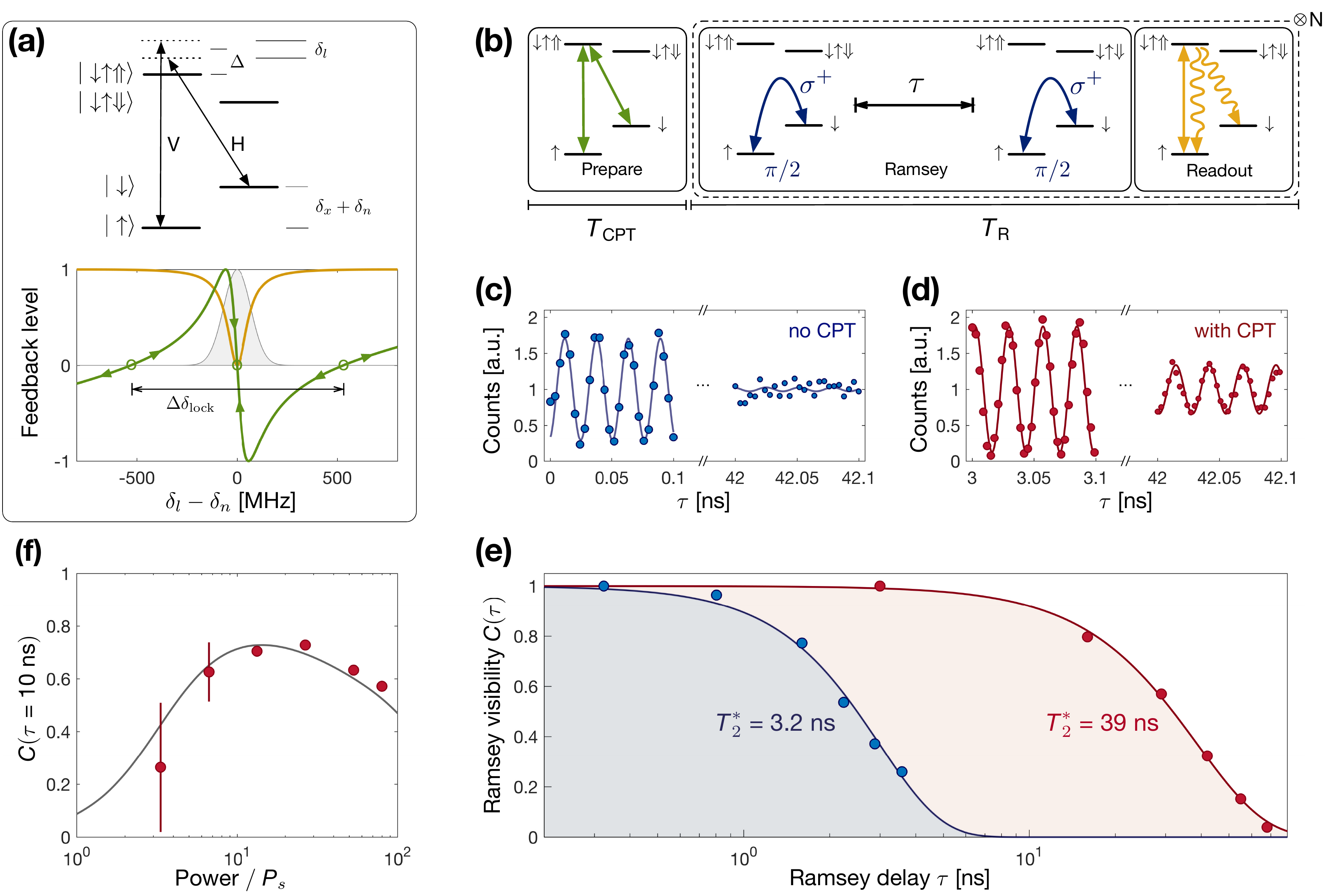}
 	\caption{\textbf{Extension of the electron $T_2^*$ via optical preparation of the nuclear ensemble}. \textbf{a} Top panel: energy levels of a singly charged QD in Voigt geometry, driven by two $\sim 965$-nm lasers with single-photon detuning $\Delta > 0$ \cite{SuppMat} from the excited trion state $\downarrow\uparrow\Uparrow$ , and two-photon detuning $\delta_l$ (H and V denotes the transition selection rules). The electron spin splitting between states $\uparrow$ and $\downarrow$ is due to the sum of the Zeeman splitting $\delta_x$ and the Overhauser shift $\delta_{n}$. Bottom panel: feedback level set by the spin polarisation in the ground state (green) and normalized scattering rate from the excited state $\Gamma_h$ around the dark-state resonance as a function of $\delta_l - \delta_{n}$ (yellow). $\Delta\delta_{\mathrm{lock}}$ represents the locking range of the feedback mechanism. The Overhauser shift probability distribution from an unprepared nuclear ensemble is also shown in grey.  \textbf{b} Experimental pulse sequence. The nuclear ensemble is prepared by driving the transitions illustrated by green arrows for a time $T_\text{CPT}$. Then, $N$ consecutive Ramsey sequences are performed during a total time $T_\text{R}$: a single sequence is composed of two circularly polarized $\pi/2$ rotation pulses separated by a delay $\tau$, followed by spin readout performed by driving the high-energy transition and measuring the resonance fluorescence. \textbf{c} Ramsey fringes measured with QD$_\text{A}$ for an unprepared and \textbf{d} prepared nuclear ensemble at 5~T with $T_\text{CPT}=840~\mu$s and $T_R=210~\mu$s. \textbf{e} Normalized Ramsey visibility as a function of the delay $\tau$ for an unprepared (blue) and prepared (red) bath. Solid curves are fits to the model $C(\tau)=\exp\left(-(\tau/T_2^*\right)^\alpha)$, where $T_2^*=3.2\pm0.1$~ns and $\alpha=2.08 \pm 0.04$ in the unprepared case and $T_2^*=39\pm 2$~ns and $\alpha=1.9\pm0.1$ in the prepared case. \textbf{f} Power dependence (relative to the saturation power of a single transition $P_s$) of the Ramsey visibility at a fixed delay $\tau = 10$~ns as measured with QD$_\text{B}$. The solid curve is calculated from a numerical simulation using a Fokker-Planck formalism.}
 	\label{Fig1}
 	\end{center}
 \end{figure*}

A spectrally sharp version of such stabilizing feedback is achieved through coherent population trapping (CPT), when driving the $\Lambda$ system formed by the two electron spin states and an excited trion state of a negatively charged QD \cite{xu2009,Onur2014,Onur2016}, as depicted in Fig.~\ref{Fig1}a. Deviations from the dark-state resonance lead to a preferential driving of one of the two optical transitions, inducing an electron spin polarisation that pulls the Overhauser field back towards a lock point set by the two-photon resonance (Fig. \ref{Fig1}a, bottom panel). The narrow spectral feature defined by the electronic dark-state coherence thereby carves out a reduced variance Overhauser field distribution from the initial thermal state with the prospect of improved qubit coherence, as inferred from a number of experiments \cite{xu2009,Sun2012,Chow2016a}. However, neither the direct measurement of such a distribution nor of its effect on the electron spin coherence has been achieved to date. In this Letter, we first prepare optically a reduced-entropy state of the QD nuclear ensemble using CPT-based feedback, and then follow its evolution as it interacts with an electron spin in the absence of feedback. We access the dephasing time, $T_2^*$, of the qubit through Ramsey interferometry with negligible perturbation to the prepared nuclear state. In this way, we demonstrate that $T_2^*$ is increased by over an order of magnitude. Further, using the qubit coherence as a probe, we observe the emergence and decay of correlations within this tailored nuclear ensemble. 

Figure~\ref{Fig1}b displays the experimental sequence used throughout this work. The nuclear ensemble is first prepared by driving the $\Lambda$ system for a time $T_{\mathrm{CPT}}$, followed by $N\sim 100$ consecutive Ramsey interference measurements on the electron spin at a fixed delay of $\tau$ performed during a time $T_{\mathrm{R}}$. Figure~\ref{Fig1}c presents the Ramsey signal measured as a function of $\tau$ in the absence of CPT preparation. There is no fringe visibility at a delay $\tau = 42$~ns indicating a complete loss of coherence. By contrast, Fig.~\ref{Fig1}d displays the Ramsey visibility for the same timescales following CPT preparation. We observe that the Ramsey fringe visibility is still significant at $\tau=42$~ns, directly showing a large extension in spin dephasing time due to ensemble preparation. 
The dephasing time increases by an order of magnitude from $3.2\pm0.1$~ns to $39\pm 2$~ns after CPT feedback (Fig.~\ref{Fig1}e), which unambiguously demonstrates the narrowing of the nuclear spin distribution.
This extension corresponds to reducing the variance of the Overhauser field by $\sim 100$; as a loose comparison, this could only be achieved with net ensemble polarisation exceeding 99\% \cite{Klauser2006}.

Qubit coherence is maximal when the width of the dark-state resonance matches the Overhauser field fluctuations of the unprepared nuclear spins, calculated from the corresponding electron $T_2^*$ to be $\Delta\delta_{\mathrm{OH}}=160\pm12$~MHz, as represented in Fig.~\ref{Fig1}a. The dependence of the Ramsey fringe visibility (at a fixed delay $\tau=10$~ns) on the optical power shown in Fig.~\ref{Fig1}f indeed reveals the optimal dark-state width to be $\Delta_\mathrm{CPT}=163\pm19$~MHz. Lower driving power, corresponding to a narrow dark-state resonance, limits the fraction of nuclear states within the locking range $\Delta\delta_{\mathrm{lock}}$ of the feedback mechanism, whereas higher driving power causes power broadening of the dark-state resonance reducing the strength of the feedback.  


\begin{figure*}[hbpt]
	\centering	
	\includegraphics[width=15cm]{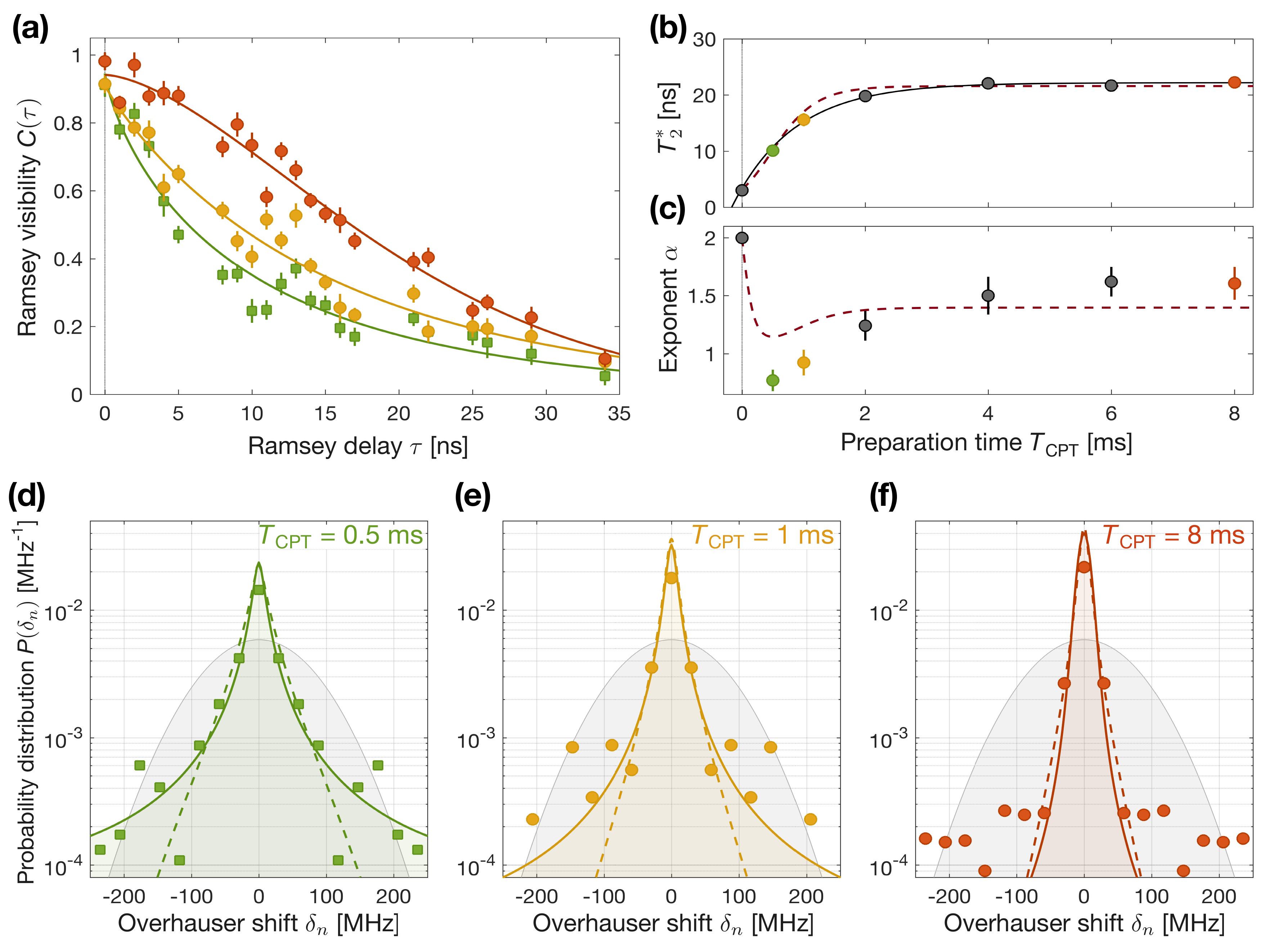}
	\caption{\textbf{Emergence of correlations within the nuclear ensemble.} These data were taken on QD$_\text{B}$, which has a lower $T_2^*$. \textbf{a} Electron Ramsey visibility for a preparation time $T_\mathrm{CPT}=0.5$~ms (green), $T_\mathrm{CPT}=1$~ms (yellow), and $T_\mathrm{CPT}=8$~ms (red). Solid curves are fit to the experimental data with $C(\tau)=A\exp\left(-(\tau/T_2^*\right)^\alpha)$, where $A$ accounts for power imbalance between the two rotation pulses. \textbf{b} Extracted $T_2^*$ and \textbf{c} $\alpha$ from the electron Ramsey visibility as a function of the preparation time. The solid curve is a phenomenological exponential fit to the data with characteristic time $T_p=0.8\pm0.2$~ms, while the dashed curves are numerical simulations. Error bars indicate a 67\% confidence interval on the fitted values. \textbf{d-f} Fourier transforms of the electron FID (symbols) and their fits (solid curves) from \textbf{a}, and simulated probability distributions of the Overhauser field (dashed curves), for $T_\mathrm{CPT}=0.5$~ms (d), $T_\mathrm{CPT}=1$~ms (e), and $T_\mathrm{CPT}=8$~ms (f). The shaded regions illustrate the Gaussian probability distribution of the unprepared nuclear ensemble. The discrepancy between the Fourier transform of the fits and experimental data below $3\times10^{-4}~$MHz$^{-1}$ is due to high-frequency noise in the experimental data.}
	\label{Fig2}
\end{figure*}

Results presented in Fig.~\ref{Fig1} have important implications for quantum information processing (QIP). The full tenfold extension of the electron dephasing time requires a preparation duty cycle $T_{\mathrm{CPT}}/(T_{\mathrm{CPT}}+T_{\mathrm{R}}) \gtrsim 40\%$ \cite{SuppMat}. Under the 1~kHz repetition rate of our experimental sequence, more than 600 Ramsey sequences or other quantum operation of $1~\mu$s duration could be performed following nuclear preparation. Moreover, the extension of $T_2^*$ well beyond the trion radiative lifetime of $\approx 0.7$~ns suppresses a key decoherence mechanism limiting the quality of spin-photon entanglement \cite{DeGreve2012,Gao2012,Schaibley2013}. Finally, changing the lock point set by the two-photon detuning provides precise control on the electron splitting to within 1.5~MHz over more than 3.5~GHz \cite{SuppMat} which can aid the generation of indistinguishable Raman photons for entanglement distribution \cite{Delteil2015,Stockill2017}.

\begin{figure*}[hbpt]
	\centering	
	\includegraphics[width=12cm]{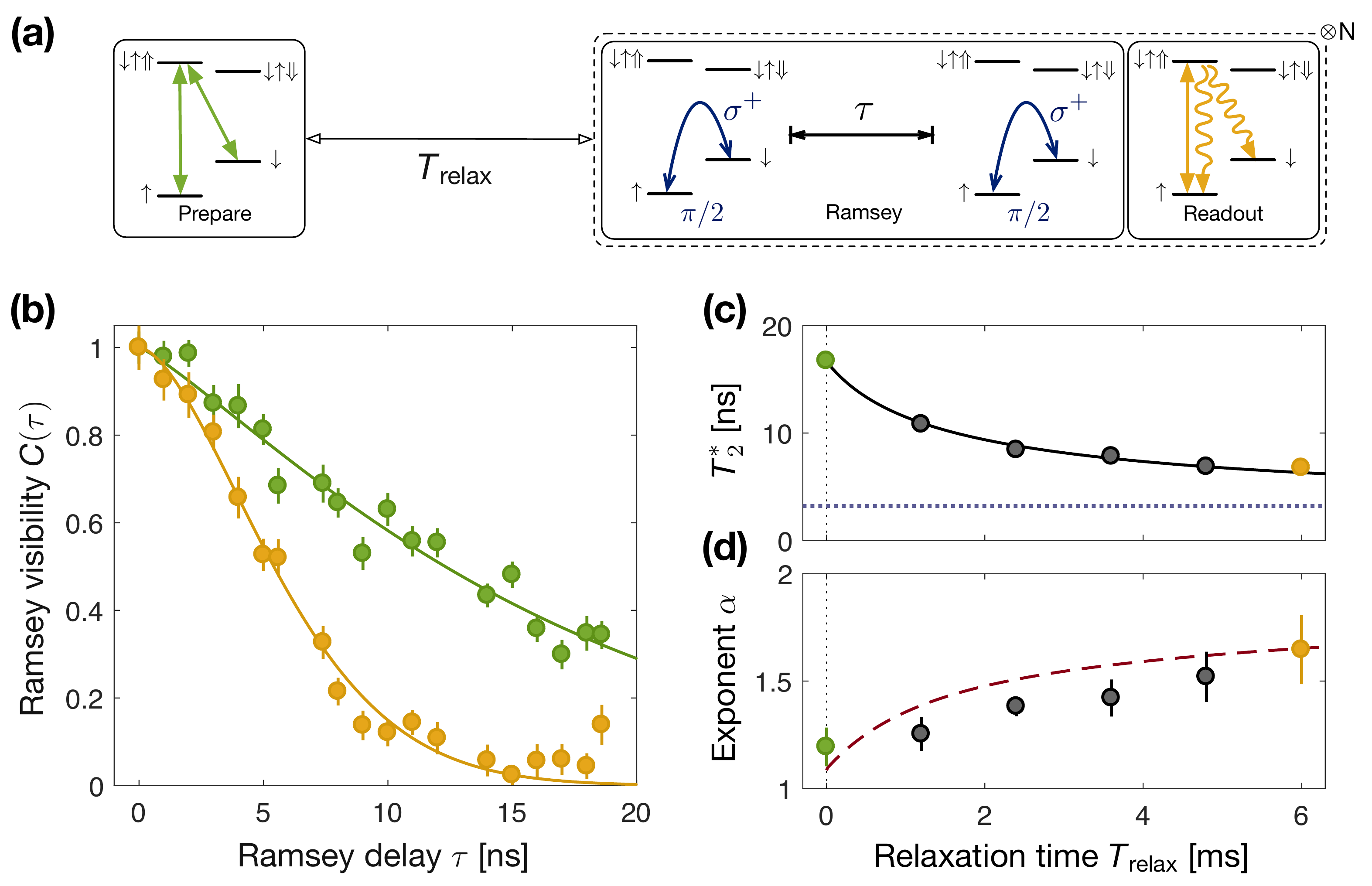}
	\caption{\textbf{Relaxation of the correlated nuclear ensemble. }\textbf{a} Pulse sequence implemented to measure the relaxation of the narrow nuclear bath. A wait time $T_{\mathrm{relax}}$ is introduced between the CPT preparation ($T_\text{CPT}=2$~ms) and the electron coherence measurement. \textbf{b} Electron FID profile for $T_{\mathrm{Relax}}=0$~ms (green) and $6$~ms (yellow). The solid curves are fitted with $C(\tau)=A\exp\left(-(\tau/T_2^*\right)^\alpha)$. \textbf{c} $T_2^*$  and \textbf{d} $\alpha$ extracted from the fits. The solid curve is fitted with $T_2^*(T_\mathrm{relax}) = T_2^*(\infty)/\sqrt{1-B\exp(-2T_\mathrm{relax}/T_c)}$, where $T_2^*(\infty)$ is the unprepared $T_2^*$ (blue dashed line in \textbf{c}), and $B = 1 - (T_2^*(\infty)/T_2^*(0))^2$, from which a correlation time $T_c=46.4\pm3.4$~ms is obtained. The dashed curve in \textbf{d} is calculated from the Fokker-Planck equation assuming an initial distribution with $\alpha=1.1$ and $T_2^*(0) =17$~ns. Error bars indicate a 67\% confidence interval on the fitted values.}
	\label{Fig3}
\end{figure*}

The modification of the nuclear spin distribution is a consequence of feedback-induced ensemble correlations, whose emergence is monitored using the electron FID profile as we vary the preparation time, $T_{\mathrm{CPT}}$. Figure~\ref{Fig2}a shows the electron spin coherence for $T_{\mathrm{CPT}}=0.5$, $1$, and $8$~ms.  We fit the visibility with $C(\tau)=A\exp\left(-(\tau/T_2^*\right)^\alpha)$, where throughout our analysis we describe the decay time and shape with $T_2^*$ and $\alpha$, respectively, thereby capturing the essential features linking the FID to the nuclear spin distribution. Figures~2b and c present $T_2^*$ and $\alpha$ as a function of $T_{\mathrm{CPT}}$. As expected, $T_2^*$ increases with preparation time to reach a steady-state value of $22$~ns over a characteristic time $T_p=0.8\pm0.2$~ms. The exponent evolves non-monotonously from $\alpha=2$, as expected for the initial gaussian state, dropping rapidly to values below 1 and later reaching a steady-state value of $\alpha=1.6$. This rich behavior suggests an interesting transient for the nuclear ensemble. Indeed, the Overhauser field probability distribution $P(\delta_{\mathrm{n}})$ is given by the Fourier transform of the FID profile, provided the high-frequency nuclear noise is negligible. This limit is achieved in an external magnetic field of 6~T, at which fast dynamics of the nuclear ensemble due to quadrupolar interactions are suppressed \cite{Bechtold2015,Stockill2016,SuppMat}. Figures~\ref{Fig2}d-f thus present the evolution of $P(\delta_{\mathrm{n}})$ corresponding to the data and the fits shown in Fig.~\ref{Fig2}a. The clear decrease in the width of the distribution is accompanied by an evolution of its shape from resembling a Lorentzian, with significant spectral weight in its wings, to resembling a Gaussian. This behavior is a direct consequence of the CPT feedback mechanism, whose Overhauser-field dependent gain imprints a transient distribution on the nuclear ensemble.


We can paint a simple picture of how the CPT-based feedback shepherds the nuclear spins into their steady-state distribution. The evolution of $P(\delta_{\mathrm{n}})$ is given by the spectral dependence of the average spin $\langle S_x(\delta_\mathrm{n})\rangle$: as can be seen in Fig.\ref{Fig1}a, there exist two Overhauser fields for which the spin imbalance is maximal and hence the feedback is the strongest. At the early stages of preparation, the probability of finding the Overhauser field close to these maximum feedback points is depleted rapidly and redistributed towards the lock point. The wings of the distribution, where the feedback is weaker, are initially unaffected. This explains the fast reduction of the exponent $\alpha$ (Fig.~\ref{Fig2}c) and of the width of the central part of the distribution (Fig.~\ref{Fig2}d,e). Then, as the preparation time is increased, only the wings of $P(\delta_{\mathrm{n}})$ can further contribute to narrowing (Fig.~\ref{Fig2}f) until $\alpha$ reaches its steady-state value.

Our measurements are consistent with theoretically anticipated values of $\alpha$, $T_2^*$ and $P(\delta_{\mathrm{n}})$ from a rate equation model (dashed curves in Fig.~\ref{Fig2}b-f). This model captures the effect of CPT on the electron spin polarisation, which in turn affects the average nuclear spin polarisation, causing an evolution of $P(\delta_{\mathrm{n}})$ under the Fokker-Planck formalism \cite{Danon2008,Onur2014,SuppMat}. The feedback on the probability distribution is governed by the time derivative of the Overhauser field,
\begin{equation}
\dot{\delta}_\mathrm{n}=-\Gamma_h(\delta_\mathrm{n})[\delta_\mathrm{n}-K\langle S_x(\delta_\mathrm{n})\rangle]-\Gamma_d\delta_\mathrm{n}.
\label{eq1}
\end{equation}
Here, $\Gamma_h(\delta_\mathrm{n})$ is the optically assisted nuclear flip rate whose spectral dependence follows the trion excited state population under CPT (yellow curve in Fig.~\ref{Fig1}a), and $\Gamma_d$ captures the dominant relaxation mechanism of the spin ensemble in the absence of optical excitation which is mediated here by the electron \cite{SuppMat}. With a hyperfine-dependent gain factor $K$, the narrowing mechanism is driven by the ground state spin polarisation $\langle S_x(\delta_\mathrm{n})\rangle$, which provides the necessary directionality to spin flips to lock the nuclear ensemble. Quadrupolar effects are known to dominate spin flips in InGaAs QDs, the directionality of the feedback mechanism is therefore likely provided by a phenomenon known as spin dragging \cite{Latta2009,Hogele2012}. An approximate steady-state solution of the Fokker-Planck equation is an Overhauser field distribution whose final variance is reduced by a factor $\propto K \Gamma_h(0) / \Delta_{\text{CPT}} \Gamma_d$ from its initial thermal variance \cite{SuppMat}. The model therefore predicts that the narrowing limit is determined by the interplay of the feedback strength $K/\Delta_{\text{CPT}}$ and the strength of nuclear spin diffusion $\Gamma_d/\Gamma_h(0)$.




In the absence of polarisation diffusion out of the QD \cite{Maletinsky2009}, the nuclear ensemble remains in its reduced-entropy state for a finite time before spin-spin interactions recover a thermal distribution over a correlation time $T_c$. This return towards thermal equilibrium can be monitored by introducing a wait time $T_{\mathrm{Relax}}$ between CPT preparation and Ramsey measurement, as shown in Fig.~\ref{Fig3}a. Figure~\ref{Fig3}b presents the Ramsey visibility for $T_{\mathrm{relax}}=0$ and 6~ms after a preparation time of 2~ms. Figures~\ref{Fig3}c and d show the fitted $T_2^*$ and $\alpha$ as a function of $T_{\mathrm{relax}}$. As $T_{\mathrm{relax}}$ increases, the distribution tends to a thermal gaussian shape ($\alpha$ increases), and the enhancement of electron coherence is lost ($T_2^*$ decreases). 
Assuming relaxation from a narrowed gaussian state whose variance evolves exponentially to its thermal value within a characteristic time $T_c$, the electron dephasing time follows the analytical expression $T_2^*(T_\mathrm{relax}) = T_2^*(\infty)/\sqrt{1-B \exp(-2T_\mathrm{relax}/T_c)}$, the fitting function used in Fig.~\ref{Fig3}c. While the nuclear distribution quickly broadens on a timescale of milliseconds, the extracted correlation time is $T_c=46.4\pm3.4$~ms. This correlation time is shorter than the estimated characteristic time of the polarisation loss $T_1$ \cite{SuppMat}, but significantly longer than the nuclear coherence time $T_2$ \cite{Wust2016}. The measurement of the nuclear relaxation allows us to fix the nuclear spin relaxation rate $\Gamma_d + \Gamma_h(0)$ in our model, which in addition to supporting our results in Fig.~2, reproduces the relaxation of $\alpha$ (Fig.~3d). The overall consistency of the model with our data supports our interpretation that we indeed prepare the optimal nuclear spin distributions leading to the maximum improvement in qubit coherence our technique offers.





We have shown that the interaction of a QD electron with its nuclei can be tailored to create reduced-entropy states of the nuclear ensemble. Such engineering of the electron spin environment results in a tenfold increase in qubit coherence, which will directly improve the transfer of quantum information between a single spin and a single photon in QDs. The magnitude of this enhancement is dictated by the feedback strength set by the hyperfine interaction and by nuclear spin diffusion. Further, access to such a correlated spin ensemble sets the stage for investigations of quantum many-body physics in QDs, possibly leading to ensemble quantum memories \cite{Taylor2003a,Taylor2003b}. Quantum correlations within the nuclei can be generated by the non-linear interactions \cite{Kitagawa1993} provided in our current feedback mechanism by the strong dependence of the electron spin polarisation on the total nuclear spin around the CPT lock point. As proposed for directly driven electron spin resonance \cite{Rudner2011}, a polarized nuclear ensemble locked around a dark-state resonance together with coherent manipulation would lead to ensemble spin squeezing.

We acknowledge financial support from the European Research Council ERC Consolidator grant agreement no. 617985 and the EPSRC National Quantum Technologies Programme NQIT EP/M013243/1. G.E-M. acknowledges financial support from NSERC. We thank M.J. Stanley and L. Huthmacher for fruitful discussions.


\end{document}